\documentclass[conference]{IEEEtran}
\IEEEoverridecommandlockouts

\usepackage{graphicx}
\usepackage{amsfonts}
\usepackage{amsmath}
\usepackage{amssymb}
\usepackage{color}
\usepackage{graphicx,color,overpic}
\usepackage{psfrag}
\usepackage{amssymb}
\usepackage{times}
\usepackage{latexsym}
\usepackage{bm}
\usepackage{cases}
\usepackage{array}
\usepackage{setspace}
\usepackage{subfigure}
\usepackage{url}
\usepackage{multirow}
\usepackage{epstopdf}

\usepackage{epsfig}
\usepackage{fancybox}
\usepackage{textcomp}

\usepackage{fancyhdr}

\ifCLASSINFOpdf
\else
\fi

\renewcommand{\vec}[1]{\mathtt{#1}}

\newcommand{\U}[0]{\vec{u}}

\newcommand{\f}[0]{\mathbf{f}}

\newcommand{\g}[0]{\mathbf{g}}
\newcommand{\x}[0]{\mathbf{x}}

\newcommand{\m}[0]{\mathbf{m}}
\newcommand{\SIM}[0]{\mathrm{sim}}

\newcommand*\Bell{\ensuremath{\boldsymbol\ell}}

\begin{document}
\title{Indoor Positioning using Similarity-based Sequence and Dead Reckoning without Training}


%\author{\IEEEauthorblockN{Ran Liu, Chau Yuen, Tri-Nhut Do, and U-Xuan Tan}
%\IEEEauthorblockA{
%Engineering Product Development Pillar\\
%Singapore University of Technology and Design\\
%8 Somapah Rd, Singapore, 487372 \\
%Email: \{ran\_liu, yuenchau, uxuan\_tan\}@sutd.edu.sg
%}
%\and
%\IEEEauthorblockN{Ye Jiang and Xiang Liu}
%\IEEEauthorblockA{School of Software and Microelectronics\\
%Peking University\\
%Daxing Dsitrict, Beijing, China, 102600\\
%Email: jiangye\_smile@pku.edu.cn, xliu@ss.pku.edu.cn}}

\author{Ran Liu, Chau Yuen, Tri-Nhut Do, Ye Jiang, Xiang Liu, and U-Xuan Tan \\
\thanks{This work is supported by Indoor Relative Positioning System project from Temasek Lab (No. IGDST1302024) and National Science Foundation of China (No. 61550110244, 61601381, and 61471306).} 
\thanks{R. Liu, C. Yuen, T. N. Do, and U-X. Tan are with the Engineering Product Development Pillar, Singapore University of Technology and Design, 8 Somapah Rd, Singapore, 487372 
{\{\tt\small ran\_liu, yuenchau, trinhut\_do, uxuan\_tan\}@sutd.edu.sg}.}
%R. Liu is also with the School of Information Engineering, Southwest University of Science and Technology, Mianyang, China, 621010.
\thanks{Y. Jiang and X. Liu are with the School of Software and Microelectronics, Peking University, Beijing, China, 102600 {\tt\small jiangye\_smile@pku.edu.cn, xliu@ss.pku.edu.cn}.
}}


%Telephone: (800) 555--1212, Fax: (888) 555--1212}


\maketitle

\thispagestyle{fancy}
\fancyhead{}
\lhead{}
\lfoot{978-1-5090-3009-5/17/\$31.00~\copyright~2017 IEEE}
\cfoot{}
\rfoot{}

\begin{abstract}
%approach to localize objects in indoor environments using a sequence of measurements from various categories of transmitters (e.g. AP and FM).
For the traditional fingerprinting-based positioning approach, 
it is essential to collect measurements at known locations as reference fingerprints during a training phase, which can be time-consuming and labor-intensive.
%The traditional fingerprinting-based positioning approach usually requires a laborious training phase to collect the measurements in an environment as the reference for positioning, 
%which is a challenge for applications involving large buildings. 
%In addition, the traditional fingerprinting method is not accurate in crowded buildings, where many people are moving and interfering with the wireless signal. 
This paper proposes a novel approach to track a user in an indoor environment by integrating similarity-based sequence and dead reckoning.
%The proposed system needs no training and is based on a smart phone only, without using any external device.
%This paper utilize location sequences as a representation of the fingerprinting map based on the geometrical relationships of the transmitters whose positions are known.
In particular, we represent the fingerprinting map as location sequences based on distance ranking of the APs (access points) whose positions are known.
%The reference fingerprinting map consists of location sequences, which are computed based on the positions of APs which are known in advance.
%by ranking Wifi access points (APs) based on their distances.
%The positions of these APs are known and can be obtained offline. 
The fingerprint used for online positioning is represented by a ranked sequence of APs based on the measured Received Signal Strength (RSS), 
which is refereed to as RSS sequence in this paper. 
%The similarity between this sequence and the fingerprint map is computed to constrcut the similarity map for the localization.
Embedded into a particle filter, we achieve the tracking of a mobile user by fusing the sequence-based similarity and dead reckoning.
%The similarities between the RSS sequence and location sequences are computed 
%and integrated with a dead reckoning from phone IMU using a particle filter to track mobile users.
%The step counting and orientation information from a phone IMU (Inertial Measurement Unit) is then integrated into particle filters to track mobile users. 
Extensive experiments are conducted to evaluate the proposed approach.
%The FM and AP have almost same localization accuracy.
\end{abstract}


% \fi
\IEEEpeerreviewmaketitle

%%%%%%%%%%%%%%%%%%%%%%%%%%%%%%%%%%%%%%%%%%%%%%%%%%%%%%%%%%%%%%%%%%%%%%%%%%%%%%%%%%%%%%%%%%%%%%%%%%%%%%%%
\section{Introduction}
\label{Introduction}
%%%%%%%%%%%%%%%%%%%%%%%%%%%%%%%%%%%%%%%%%%%%%%%%%%%%%%%%%%%%%%%%%%%%%%%%%%%%%%%%%%%%%%%%%%%%%%%%%%%%%%%%
%Location-based services have shown an increasing interest
Research community has shown an increasing interest in indoor positioning due to the rapid demand of location-based services\,\cite{hasalaidentifying}. 
In the literature, various techniques including Received signal strength (RSS)\,\cite{Liu_trackingRFID_2015}, time-of-arrival (TOA)\,\cite{ran_ICRA2017_cooperative}, and angle-of-arrival (AOA)\,\cite{aoa_monopole_ieee_sensor_2015} 
have been used for positioning.
%A majority of indoor localization systems employ RSS since it is readily to be extracted in many off-the-shelf devices such as Wifi, RFID, and Bluetooth.
%The existing WLAN-based infrastructures are usually covered by a number of wireless access points (APs), which can provide the receive signal strength to infer mobile's locations.
A number of propagation model-based or fingerprinting-based techniques have been proposed\,\cite{Yassin_ieee_tutorials_2016}. 

Propagation model-based approach\,\cite{ElbaklyY16} needs a model to explicitly characterize the propagation of radio signals.
Its accuracy is limited due to multipath issues of radio signal propagation in indoor environments.
On the contrary, fingerprinting-based approach\,\cite{koch2016icra} represents locations using a priori sets of sensor measurements collected during an offline training phase. 
The location of a user is then determined by matching current measurement with reference fingerprints.
These approaches are shown to have better accuracy as compared to the model-based approaches.

%An approach to reduce the complexity of collecting the measurements is to use a kind of path-loss model to generate the offline fingerprinting map.
%However, the accuracy and complexity need to be investigated.

\begin{figure}
  \centering
    \subfigure[Reference fingerprinting map]{
\label{fig:fingerprinting_map}
    \includegraphics[height=0.15\textwidth]{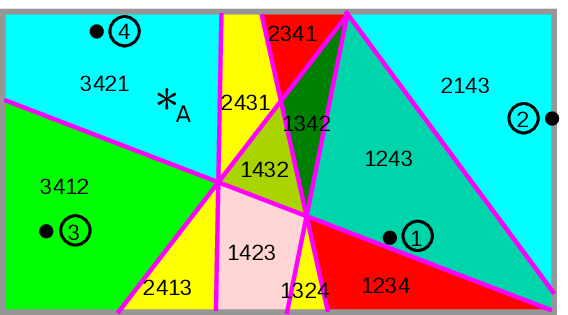}
    }   
    \hspace{-0.45cm}
  \subfigure[Location sequence at point $A$]{
\label{fig:example_location_sequence}
        \includegraphics[height=0.15\textwidth]{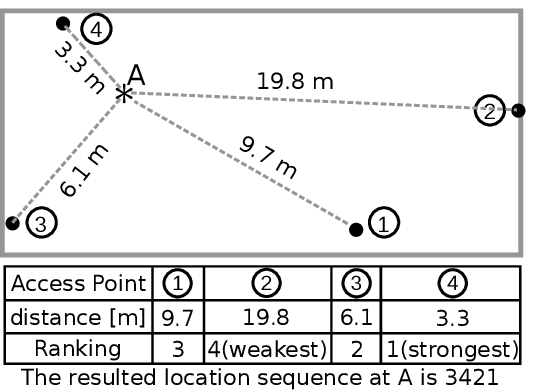}
        }
   \caption[Location sequence]
{(a) A fingerprinting map constructed by four access points which are denoted by dark dots. 
The sequence inside each region (i.e., $3412$ or $2413$) represents the location sequence (reference sequence) of this region.
(b) Location sequence of position $A$ based on the ranking of the distance away from the access points.
}
\label{fig:example_basci_idea}
\vspace{-0.5cm}
\end{figure}

%\begin{figure}
%\centering
%\includegraphics[width=0.4\textwidth]{images/1234}
%\caption{Example of a fingerprinting map representing the environment with four transmitters. 
%Dark dots denote the position of the transmitters. 
%\textcircled{n} denotes the label of the transmitter. 
%The sequence inside each region (for example $3412$ or $2413$) represents the location sequence (reference sequence) of this region, which is obtained by the distance away from the transmitters. }
%\label{fig:fingerprinting_map}
%\vspace{-0.1in}
%\end{figure}

A good positioning accuracy is guaranteed by a time-consuming and thorough site survey phase which collects the radio measurements at reference locations through the environment.
%Moreover, the complexity of the fingerprinting-based approaches is proportional to the number of entities one wants to locate.
%For example, in order to locate a HP ProBook laptop and a Samsung galaxy S5 smart phone, one needs to build up a fingerprint map for each of them.
%\textcolor{red}{//we have to write some sentences about this.}
Although different techniques\,\cite{Yang_using_human_motions}\,\cite{Zero_calibration} are proposed to reduce this phase, 
maintaining the fingerprinting map is still labor-intensive due to the change of the environment.

To overcome the tedious site survey phase to construct the fingerprinting map, 
we use the sequence-based approach, which is based on our previous work in \,\cite{Liu_Relative_Globecom}\,\cite{Liu_selective_ap_vtc}.
The technique is also used by other researchers\,\cite{tracking_unreliable_node_seq}\,\cite{knn_sequence_2016} to localize and track mobile users. 
%Yedavalli {et al.}\cite{Ecolocation_extended} introduced sequence-based technique for indoor localization. 
%For example, Zhong \textit{et al.}\,\cite{tracking_unreliable_node_seq} utilized RSS sequence to track mobile users by optimal path matching. 
%In our previous work\,\cite{Liu_Relative_Globecom}, we used the sequence-based approach to fuse Wifi RSS with UWB ranging measurement for relative positioning.
%labrious these problems constrcut  only knowing the positions of the transmitters. 
In particular, the sequence-based fingerprinting map consists of a set of connecting regions, 
which is represented by a ranked sequence (i.e., location sequence) based on the distance to the APs. 
%For the rest of this paper, this sequence is referred to as the location sequence, since this sequence is supposed to uniquely represent the location of a region. 
Fig.\,\ref{fig:fingerprinting_map} shows one example of the fingerprinting map.
%To differentiate our approach from the fingerprinting-based approach, our approach is referred to as the sequence-based approach. 
In the online localization and tracking phase, 
the RSS sequence is formulated by ranking the measured RSS of the APs in descending order. 
The location of a user can be simply determined by those location sequences whose similarities best match the RSS sequence\,\cite{Liu_Relative_Globecom}.
%and the location sequence in the reference fingerprinting map\,\cite{Liu_Relative_Globecom}. 
%The key novelty of this paper is to compare the similarity between the RSS sequence and the location sequence stored in fingerprint map in order to determine the position of a user.

%To address these challenges and achieve a better localization accuracy, 
%we propose to fuse the dead reckoning from low-cost inertial-measurement unit (IMU) sensors.  
%which can provide a good displacement estimation over a short period of time by dead reckoning. 

%The reported accuracy of this approach is reported to be 3 and 5 meters depending on the number.
%We can achieve a good position estimation over short periods of time using the low-cost inertial-measurement unit (IMU) sensors by dead reckoning, 
%but IMU is notorious for accumulated error over long term run. 
Due to multipath effect on radio signal propagation, 
it is common that the measured RSS sequence does not match the true location sequence, thus resulting in a poor positioning accuracy. 
%Dead reckoning achieves the tracking of positions by integrating the measurements from IMU (inertial measurement unit), which is accurate for a short period of time. 
Due to cost-effective feature, modern smart phones are equipped with IMU (inertial measurement unit) sensors.
These sensors can be used to implement a dead reckoning which can precisely track the position of a user for short period of time. 
%smart phone are very good at position estimation
However, the error is accumulated for long term run, which must be corrected by other sources of sensors. 
Therefore, we propose a novel method to track a user by fusing similarity-based sequence and dead reckoning using a particle filter. 
The proposed approach can incorporate the measurements from various sources of sensors (e.g., Wifi and IMU) 
with complementary error characteristics to improve the positioning accuracy. 
Moreover, our approach does not require the tedious training phase to construct the fingerprinting map,
as compared to the traditional fingerprinting-based approach.
%the  do not need any training time to construct our reference fingerprinting map can be ignored, as compared to the tedious site survey in the traditional fingerprinting-based approach.

% To be precise, we would like to highlight features of our sequence-based approach:
We highlight the contributions of this paper as follows:
\begin{itemize}
 \item We propose to fuse similarity-based sequence based on relative signal strength for the tracking of mobile users without the need for training.
  \item We design a particle filtering that fuses Wifi and IMU measurements to achieve a better tracking accuracy. 
  %\item We compared our approach to the traditional fingerprinting-based approach to show its performance.
 \item We implemented our approach and evaluated its performance through extensive experiments. Note that the whole implementation is using a smart phone only, without any external device.
 \end{itemize}

We organize the rest of this paper as follows. 
We present the system overview in Sect.\,\ref{system_overview}, 
which is followed by the details of the particle filtering in Sect.\,\ref{particle_filtering_fusion}. 
We show the experimental details in Sect.\,\ref{experimental_evaluations} and conclude this paper in Sect.\,\ref{conclusions}.

\begin{figure}
\centering
\includegraphics[width=0.40\textwidth]{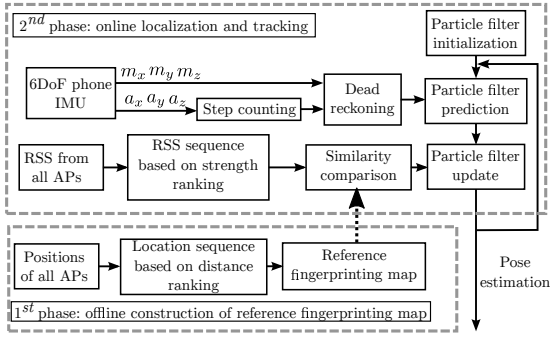}
\caption{System overview.}
\label{fig:system_overview}
\vspace{-0.5cm}
\end{figure}

%%%%%%%%%%%%%%%%%%%%%%%%%%%%%%%%%%%%%%%%%%%%%%%%%%%%%%%%%%%%%%%%%%%%%%%%%%%%%%%%%%%%%%%%%%%%%%%%%%%%%%%%
\section{System overview}
\label{system_overview}
%%%%%%%%%%%%%%%%%%%%%%%%%%%%%%%%%%%%%%%%%%%%%%%%%%%%%%%%%%%%%%%%%%%%%%%%%%%%%%%%%%%%%%%%%%%%%%%%%%%%%%%%
%In wireless-based infrastructures, various wireless sensors, for example Wifi, Bluetooth, RFID, and FM, are employed to provide different services.
%These sensors are able to report the received signal strength of devices, 
%which can be used to infer their locations. 
%The traditional fingerprinting-based approaches require a labor-intensive site survey to characterize the RF features in the environment.
This paper proposes a novel approach to combine similarity-based sequence and dead reckoning to localize and track users without the need of training.
To be precise, we use a sequence-based technique to construct the fingerprinting map without human intervention. 
%An overview of our system is shown in Fig.\,\ref{fig:system_overview}. 
As illustrated in Fig.\,\ref{fig:system_overview}, our proposed system consists of two phases, 
namely: 1) a offline phase to construct the reference fingerprinting map and 2) an online phase to localize and track mobile users. 

%The map is in essence a set of small connecting regions, obtained by intersecting the lines connecting each transmitter in the middle without laborious site survey. 
%\begin{figure}
%\centering
%\includegraphics[width=0.4\textwidth]{images/1234_distance}
%\caption{\emph{Example}: Location sequence of position $A$ based on the ranking of the distance to the transmitters.
%}
%\label{fig:example_location_sequence}
%\vspace{-0.1in}
%\end{figure}

%%%%%%%%%%%%%%%%%%%%%%%%%%%%%%%%%%%%%%%%%%%%%%%%%%%%%%%%%%%%%%%%%%%%%%%%%%%%%%%%%%%%%%%%%%%%%%%%%%%%%%%%
\subsection{Offline Construction of Reference Fingerprinting Map}
\label{Fingerprint_map_generation}
%%%%%%%%%%%%%%%%%%%%%%%%%%%%%%%%%%%%%%%%%%%%%%%%%%%%%%%%%%%%%%%%%%%%%%%%%%%%%%%%%%%%%%%%%%%%%%%%%%%%%%%%
In this phase, the reference fingerprinting map is constructed by partitioning the environment into a set of regions. 
Each region is associated with a location sequence, which is represented as the ranking of APs based on their distance in ascending order.
This phase results in a set $\m=\{(\f_1,\Bell_1),...,(\f_M,\Bell_M)\}$ of $M$ fingerprints, 
where $\f_i$ is the location sequence and $\Bell_i=(x_i, y_i)$ is the 2D location. 

An example of the fingerprinting map constructed with four APs is shown in Fig.\,\ref{fig:fingerprinting_map}, 
where \textcircled{$k$} denotes the location of the $k^{th}$ access point. 
We show an example to compute the location sequence at a position in Fig.\,\ref{fig:example_location_sequence}.
%Fig.\,\ref{fig:example_location_sequence} gives an example of the construction of one location sequence.
In this example, 
the order of reference APs is predefined as $\textcircled{1}\textcircled{2}\textcircled{3}\textcircled{4}$. 
Ranking the APs in ascending order based on their distances away from $A$, 
we will get the location sequence at location $A$: $\f=3421$.
%we fix the predefined order of reference APs as $\textcircled{1}\textcircled{2}\textcircled{3}\textcircled{4}$. 
%Ranking the APs in ascending order based on their distances to $A$, 
%we will get the location sequence at location $A$: $\f=3421$.

\begin{figure}
  \centering
    \subfigure[]{
\label{fig:wifi_measurements}
        \includegraphics[width=0.32\textwidth]{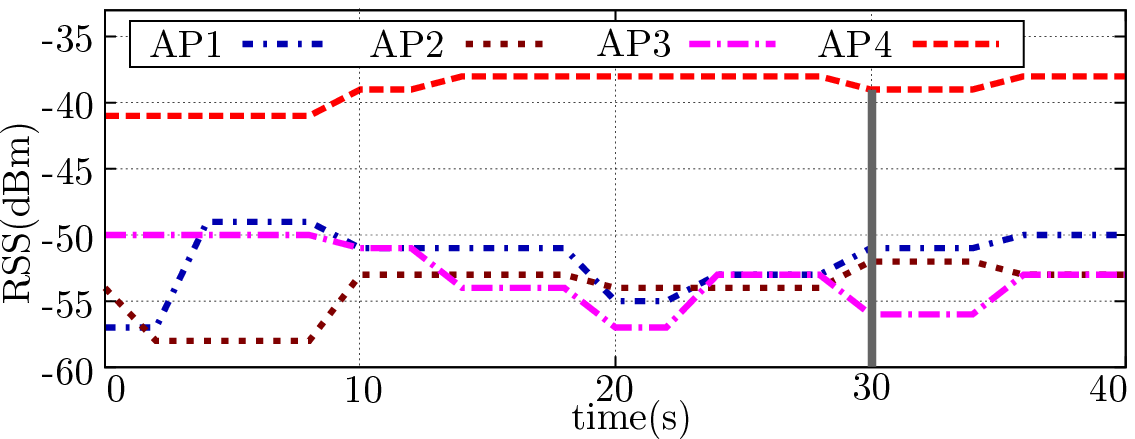}
        }        
        \hspace{-0.2cm}
       \subfigure[]{
\label{fig:fuse_signature}
        \includegraphics[width=0.14\textwidth]{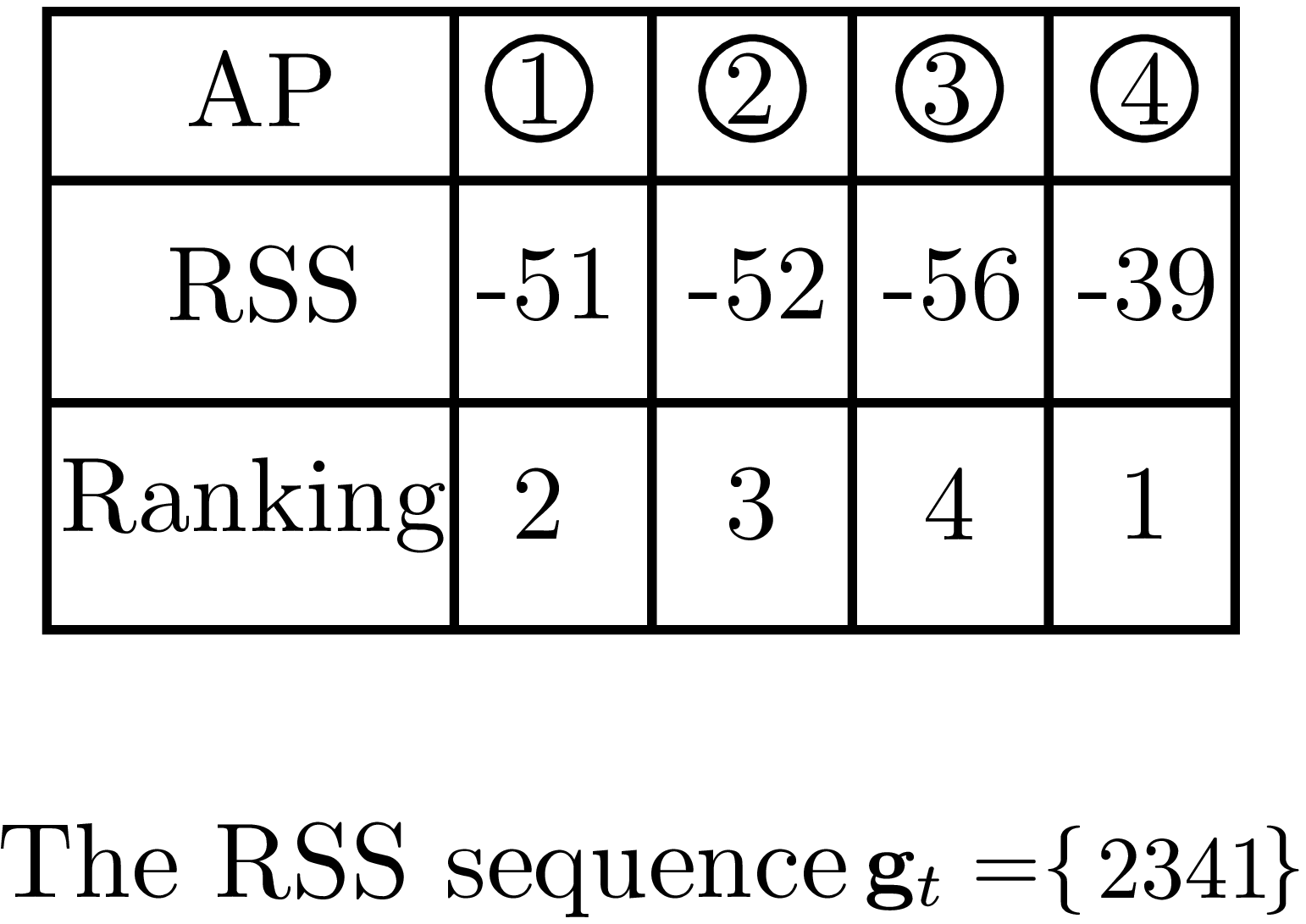}
        }

   \caption[Wifi measurements and the RSS sequence.]
{(a) The signal strength from four access points during a period of time. 
(b) The RSS sequence at $t=30$\,s.}
\label{fig:rss_values}
\vspace{-0.5cm}
\end{figure}

%We assume the locations of the APs are known in advance. 
%This is reasonable since in many commercial scenarios, the locations of the APs are fixed to optimize the performance of the wireless sensor network. 
%In case that the positions of the APs are not known, researchers proposed different techniques to localize them \cite{Zhuang_ap_localization}.  
%For example, authors in\,\cite{Zhuang_ap_localization} proposed an effective way to automatically locate the positions of APs based on an inertial navigation solution.

%This is not quite correct. 

%%%%%%%%%%%%%%%%%%%%%%%%%%%%%%%%%%%%%%%%%%%%%%%%%%%%%%%%%%%%%%%%%%%%%%%%%%%%%%%%%%%%%%%%%%%%%%%%%%%%%%%%
\subsection{Online Localization and Tracking}
\label{RSS_sequence}
%%%%%%%%%%%%%%%%%%%%%%%%%%%%%%%%%%%%%%%%%%%%%%%%%%%%%%%%%%%%%%%%%%%%%%%%%%%%%%%%%%%%%%%%%%%%%%%%%%%%%%%%
In the online phase, we measure the RSS from APs and formulate them as the RSS sequence by ranking the APs based on the strength in descending order. 
%the RSS sequence is represented by a ranked sequence of the transmitters by sorting the signal strength in descending order. 
Fig.\,\ref{fig:fuse_signature} shows an example to compute the RSS sequence at $t=30$\,s in Fig.\,\ref{fig:wifi_measurements}. 

%%%%%%%%%%%%%%%%%%%%%%%%%%%%%%%%%%%%%%%%%%%%%%%%%%%%%%%%%%%%%%%%%%%%%%%%%%%%%%%%%%%%%%%%%%%%%%%%%%%%%%%%
%\subsection{Noisy RSS measurements in uncontrolled environment}
%\label{nosiy_rss}
%%%%%%%%%%%%%%%%%%%%%%%%%%%%%%%%%%%%%%%%%%%%%%%%%%%%%%%%%%%%%%%%%%%%%%%%%%%%%%%%%%%%%%%%%%%%%%%%%%%%%%%%
In theory, 
the measured RSS sequence should fully match the location sequence of the region where the user locates. 
%Assuming an indoor environment with windows, partitions, furniture, equipments, and people walking around, 
In practice, 
radio signal propagation suffers from multi-path effect mainly due to reflection surfaces in the environment.
Therefore, it is not surprised that the measured RSS sequence is not identical to the true location sequence. 
For example, Fig.\,\ref{fig:wifi_measurements} shows the signal strength from four access points at location $A$ in Fig.\ref{fig:example_basci_idea} for a duration of 40 seconds. 
The true location sequence at this location is $3421$, while the measured RSS sequence is $\g_t=2341$ at $t=30$\,s (see Fig.\,\ref{fig:rss_values} in detail). 

The measured RSS sequence $\g_t$ is then matched against the location sequence $\f_i$ in the fingerprinting map $\m$ 
to compute the similarity-based sensor model for the correction of particle filtering (see Sect.\,\ref{state_estimation}). 
We further integrate step counting and orientation information from IMU into a particle filter to track a mobile user. 
An overview of the online localization and tracking using a particle filter with similarity comparison is shown in Fig.\,\ref{fig:system_overview} 
and will be described in the next section.

%%%%%%%%%%%%%%%%%%%%%%%%%%%%%%%%%%%%%%%%%%%%%%%%%%%%%%%%%%%%%%%%%%%%%%%%%%%%%%%%%%%%%%%%%%%%%%%%%%%%%%%%
\section{Particle Filtering with Similarity Comparison}
\label{particle_filtering_fusion}
%%%%%%%%%%%%%%%%%%%%%%%%%%%%%%%%%%%%%%%%%%%%%%%%%%%%%%%%%%%%%%%%%%%%%%%%%%%%%%%%%%%%%%%%%%%%%%%%%%%%%%%%
%An overview of the sensor fusion using a particle filter is shown in Fig.\,\ref{fig:fusion_overview}. 
%%%%%%%%%%%%%%%%%%%%%%%%%%%%%%%%%%%%%%%%%%%%%%%%%%%%%%%%%%%%%%%%%%%%%%%%%%%%%%%%%%%%%%%%%%%%%%%%%%%%%%%%
%\subsection{Online Localization Using a Particle Filter}
\subsection{Particle Filtering}
\label{state_estimation}
%%%%%%%%%%%%%%%%%%%%%%%%%%%%%%%%%%%%%%%%%%%%%%%%%%%%%%%%%%%%%%%%%%%%%%%%%%%%%%%%%%%%%%%%%%%%%%%%%%%%%%%%
%State estimation from noisy sensor measurements is a key step towards implementing a reliable and real-time tracking system. 
%In the context of our indoor positioning, 
%state estimation addresses the problem of inferring the pose of a user by fusing RSS measurements from Wifi with IMU data. 
We consider the estimation of the pose of user $\x_{t}$ at time $t$ as Bayesian inference. 
Formally, we denote $\g_{1:t}$ as the Wifi measurements until time $t$, 
$\U_{t}$ as the dead reckoning input from IMU sensor,
and $\m$ as the reference fingerprinting map.
The goal is to estimate the posterior probability $p(\x_{t}|\g_{1:t},\m,\U_{1:t})$.
Based on Bayesian inference, we can further factorize $p(\x_{t}|\g_{1:t},\U_{1:t},\m)$ into:
%to estimate the pose $\x_{t}$ of a user at time $t$, 
%we need to know the posterior probability $p(\x_{t}|\g_{1:t},\m,\U_{1:t})$.
%Here $\x_{t}$ is the pose of a user to be tracked at time $t$, 
%$\g_{1:t}$ are the Wifi measurements until time $t$, 
%$\m$ is the reference fingerprinting map, 
%and $\U_{t}$ is the dead reckoning input from IMU sensor. 
%Based on Bayesian inference, we can factorize $p(\x_{t}|\g_{1:t},\m,\U_{1:t})$ into:
\begin{equation}
 \begin{split}
  p(\x_{t}&|\g_{1:t},\U_{1:t},\m) =\eta_t \cdot  p(\x_{t}|\x_{t-1},\U_t)\\
& \cdot { p(\g_t|\x_{t},\m) } \cdot p(\x_{t-1}|\g_{1:t-1},\U_{1:t-1},\m),
 \end{split}
  \label{eq:bayesian_framework}
 \end{equation}
where $\eta_t$ is a normalizer to ensure that the sum of total probability equals to one. 
$p(\x_{t}|\x_{t-1},\U_t)$ is the motion model, 
which predicts the pose of a user at time $t$ based on the previous pose $\x_{t-1}$ and dead reckoning from IMU $\U_t$. 
${p(\g_{t}|\x_{t},\m)}$ is the observation model, which represents the likelihood of receiving a measurement $\g_t$ at pose $\x_t$ given the reference fingerprinting map $\m$ (see Sect.\,\ref{similarity_measures}). 
We choose the particle filter as an implementation due to its non-parametric feature.

For the particle filtering, the pose of a user $\x_t$ is represented by a set of particles $\x_t=\{\x_t^{[i]},w_t^{[i]}\}_{i=1}^{N}$, where $N$ is the number of particles. 
Each particle consists of pose hypotheses $\x_t^{[i]}=\{x_t^{[i]},y_t^{[i]},\theta_t^{[i]}\}$ (i.e., 2D position $\{x_t^{[i]},y_t^{[i]}\}$ and orientation $\theta_t^{[i]}$ ) and the weight $w_t^{[i]}$.
%The position of the tag is computed by a weighted mean among all particles. % $\overline{l}_{j}=(\overline{x},\overline{y})=\sum_{i=1}^{N} w^{(i)}\x^{(i)}$. 
In general, the particle filter is executed recursively with the following three steps (also shown in Fig.\,\ref{fig:system_overview}):
1) \textbf{Prediction:} draws a new set of particles according to the motion model $p(\x_{t}|\x_{t-1},\U_t)$, which is determined by the dead reckoning input of the IMU (see Sect.\,\ref{IMU} for more detail). 
2) \textbf{Correction:} assigns each particle with a new weight according to the observation model ${p(\g_{t}|\x_{t},\m)}$ when a new measurement $\g_{t}$ arrives (see Sect.\,\ref{similarity_measures}), 
i.e., $w_t = \eta_t \cdot w_{t-1} \cdot p(\g_{t}|\x_{t},\m)$. 
3) \textbf{Resampling:} generates a set of new particles as a replacement of the old set of particles based on their weights. 

%%%%%%%%%%%%%%%%%%%%%%%%%%%%%%%%%%%%%%%%%%%%%%%%%%%%%%%%%%%%%%%%%%%%%%%%%%%%%%%%%%%%%%%%%%%%%%%%%%%%%%%%
\subsection{Observation Model based on Sequence Similarity}
\label{similarity_measures}
%%%%%%%%%%%%%%%%%%%%%%%%%%%%%%%%%%%%%%%%%%%%%%%%%%%%%%%%%%%%%%%%%%%%%%%%%%%%%%%%%%%%%%%%%%%%%%%%%%%%%%%%
The observation model ${p(\g_{t}|\x_{t},\m)}$ represents the likelihood of receiving a measurement $\g_t$ at pose $\x_t$ given the location sequence map 
$\m=\{(\f_1,\Bell_1),...,(\f_M,\Bell_M)\}$.
Similar to\,\cite{rfid_path_following_iros_2012}, 
we approximate ${p(\g_{t}|\x_{t},\m)}$ using weighted $k$-nearest neighbors (WKNN) approach. 
Based on a similarity measure $\SIM(\g_t,\f_i)$, we could obtain the $k$ reference fingerprints ${\f_{\pi(1)},...,\f_{\pi(k)}}$
whose similarities best match the measured RSS sequence $\g_t$. 
Then ${p(\g_{t}|\x_{t},\m)}$ is approximated as:
\begin{equation}
\label{equ:similarity}
p(\g_t|\x_{t},\m) \approx \sum_{j=1}^{k}{\SIM(\g_t,\f_{\pi(j)})\exp(-\frac{1}{2}d^{2}(\x_t,\Bell_{\pi(j)}))},
\end{equation}
where $d^{2}(\cdot)$ is a squared distance measure to assess the translational displacement.
\begin{equation}
d^{2}(\x_t,\Bell_{\pi(j)})=\frac{(x_t-x_{\pi(j)})^2}{\lambda}+\frac{(y_t-y_{\pi(j)})^2}{\lambda},
\end{equation}
where $\lambda$ is parameter to control the bandwidth of the translational displacement. 
The impact of parameter $\lambda$ on the tracking accuracy is shown in Sect.\,\ref{sect:partilce_size}.

We use Kendall Tau coefficient to compute the similarity $\SIM(\g_t,\f_i)$ between the measured RSS sequence $\g_t$ and location sequence $\f_i$:
\begin{equation}
\label{eq:kd_coefficient}
\tau(\g_t,\f_i) = \frac{n_c(\g_t,\f_i)-n_d(\g_t,\f_i)}{\frac{1}{2}n(n-1)},
\end{equation}
where $n_c(\g_t,\f_i)$ and $n_d(\g_t,\f_i)$ are 
the numbers of concordant pairs and discordant pairs between $\g_t$ and $\f_i$ respectively and $n$ is the length of $\g_t$ and $\f_i$. 
As a requirement, the similarity usually lies in 0 and 1, 
therefore $\SIM(\g_t,\f_i) = \frac{1+\tau}{2}$.
   
 \begin{figure}
  \centering
    \subfigure[]{
\label{fig:experimental_setup}
    \includegraphics[width=0.24\textwidth]{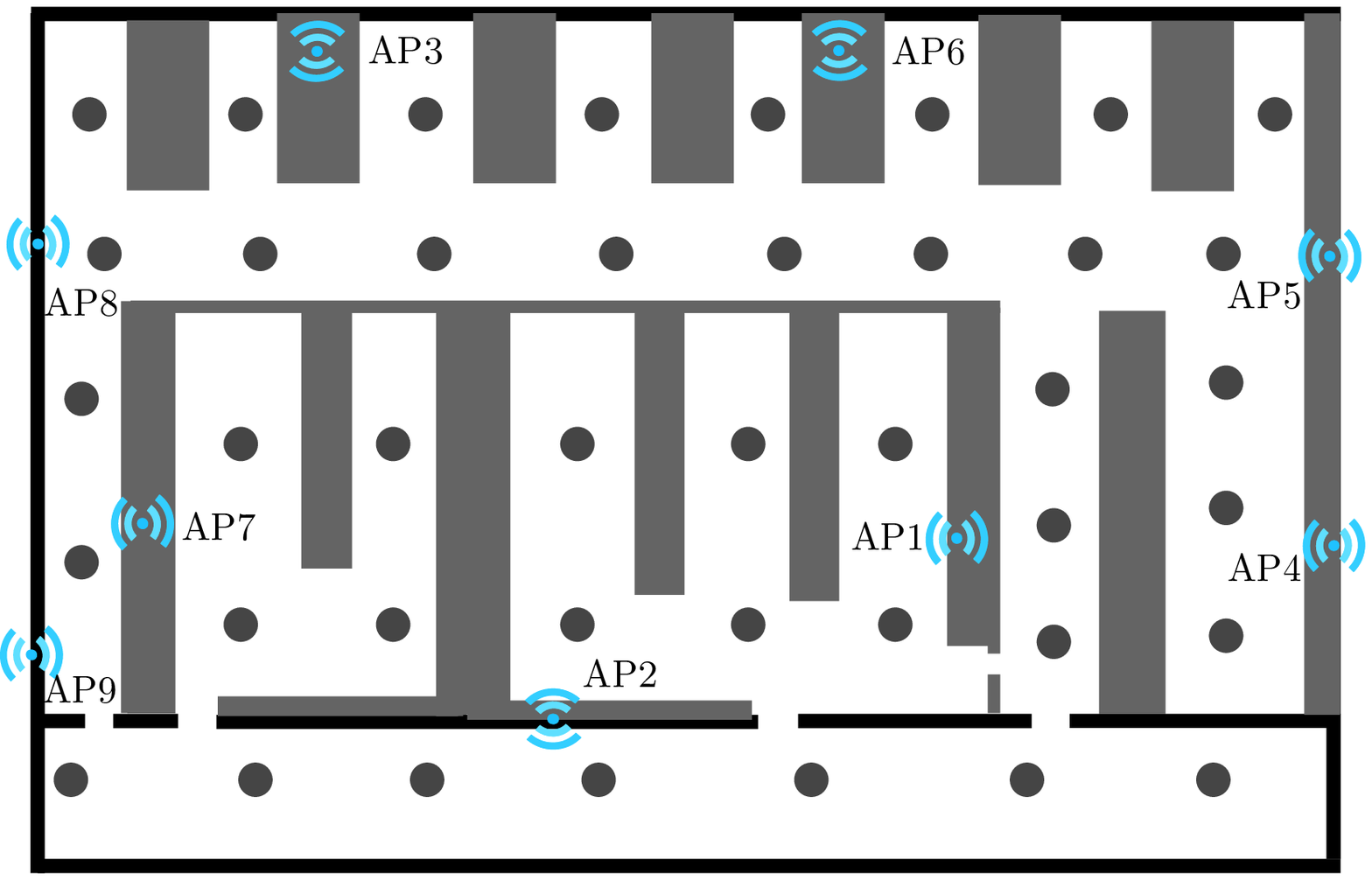}
    }    
  \subfigure[]{
\label{fig:experimental_snapshot}
        \includegraphics[width=0.20\textwidth]{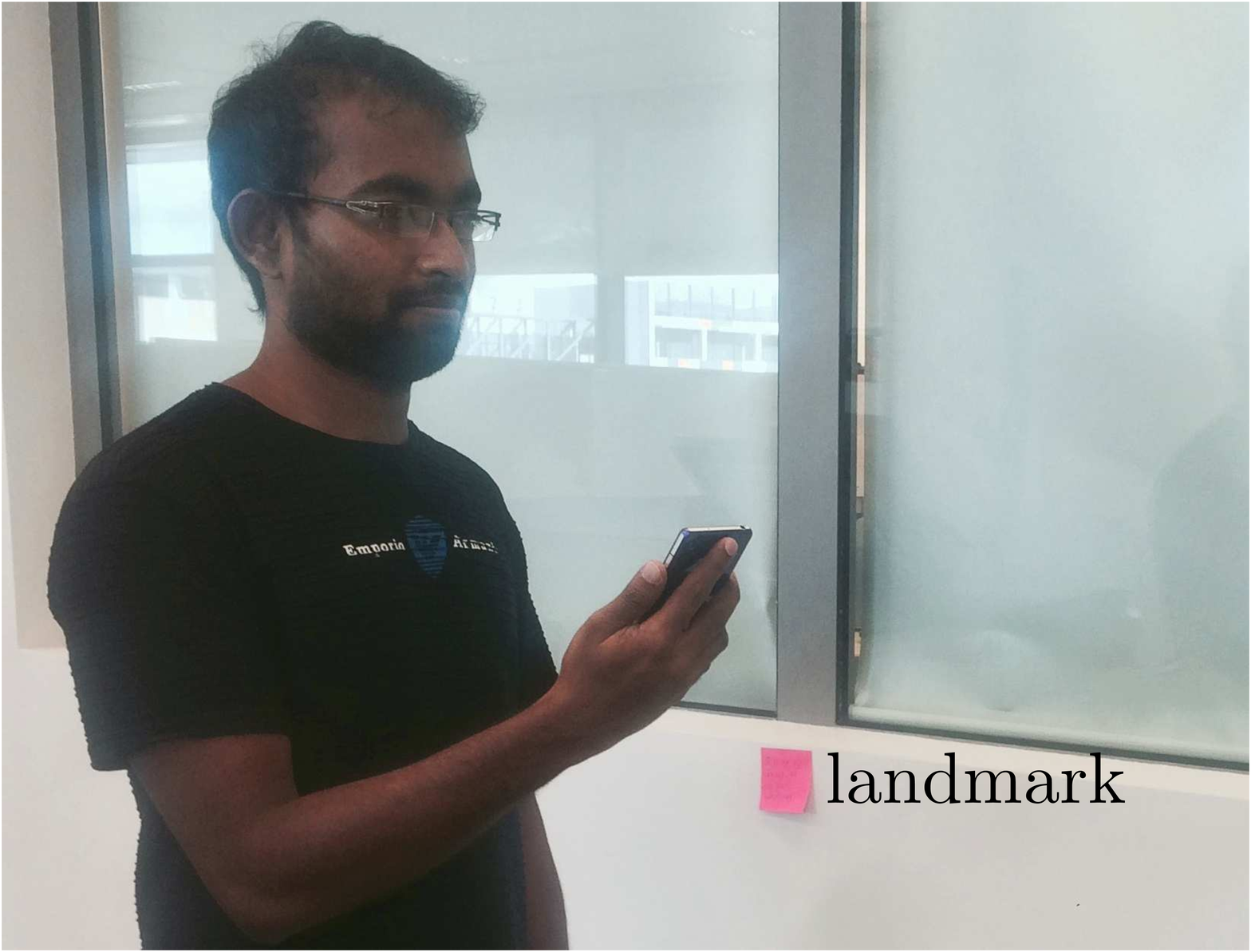}
        }
   \caption[Environmental setup and sensors carried by a user.]
{Illustration of the experimental setup. 
(a) Experimental environment and the locations (black dots) where reference fingerprints are manually collected.
(b) One experimental snapshot.
}
\label{fig:environment}
\vspace{-0.5cm}
\end{figure}

%%%%%%%%%%%%%%%%%%%%%%%%%%%%%%%%%%%%%%%%%%%%%%%%%%%%%%%%%%%%%%%%%%%%%%%%%%%%%%%%%%%%%%%%%%%%%%%%%%%%%%%%
\subsection{Fuse Dead Reckoning Information from Phone IMU}
\label{IMU}
%%%%%%%%%%%%%%%%%%%%%%%%%%%%%%%%%%%%%%%%%%%%%%%%%%%%%%%%%%%%%%%%%%%%%%%%%%%%%%%%%%%%%%%%%%%%%%%%%%%%%%%%
We utilize the IMU sensor inside the phone to achieve dead reckoning. 
%IMU uses the accelerometers, gyroscopes, and magnetometers to estimate the movement of a device. 
The IMU consists of a 3D accelerometer, a 3D gyroscope, and a 3D magnetometer. 
We implemented the auto-correlation based step counting in\,\cite{Zero_calibration}.
Given the accelerometer data, 
\cite{Zero_calibration} achieved step counting by discovering the periodic step patterns through normalized auto-correlation.
%We fix the step length $s$ in our case. 
The magnetometer reading from the IMU is used as the orientation of the user by assuming the phone is always held by a person in front of him during walking.

As a result, the phone will send the current step counting $c_t$ and the orientation $\alpha_t$ (i.e., $\U_t=(c_t,\alpha_t)$) 
to the server for the sensor fusion (see Fig.\,\ref{fig:system_overview}).
The state of a particle is predicted based on the dead reckoning corrupted with a Gaussian noise:
%We predict the state of particles upon the IMU measurement corrupted with a Gaussian noise: 
%\begin{align}
%\x_t=\x_{t-1}+\Delta\x_{t}+\sqrt{\Delta x_t^2+\Delta y_t^2} \cdot \mathcal{N}(0, \sigma^2),
 % \label{eq:motion_model}
 %\end{align}
\begin{align}
x_t=x_{t-1}+s\cdot(c_t-c_{t-1}) \cdot \cos (\theta_{t-1}) \cdot (1+\mathcal{N}(0, \sigma_d^2))\\
y_t=y_{t-1}+s\cdot(c_t-c_{t-1}) \cdot \sin (\theta_{t-1}) \cdot (1+\mathcal{N}(0, \sigma_d^2))\\
\theta_t=\theta_{t-1}+(\alpha_t-\alpha_{t-1}) \cdot (1+\mathcal{N}(0, \sigma_{\theta}^2)),
  \label{eq:motion_model}
 \end{align}
 where $s$ is the step length, $\sigma_d$ and $\sigma_\theta$ are Gaussian noises added to distance displacement and orientation respectively.
 
\begin{figure*}
  \centering    
     \subfigure[Trajectories with and without IMU]{
\label{fig:trajectory}
        \includegraphics[height=0.2105\textwidth]{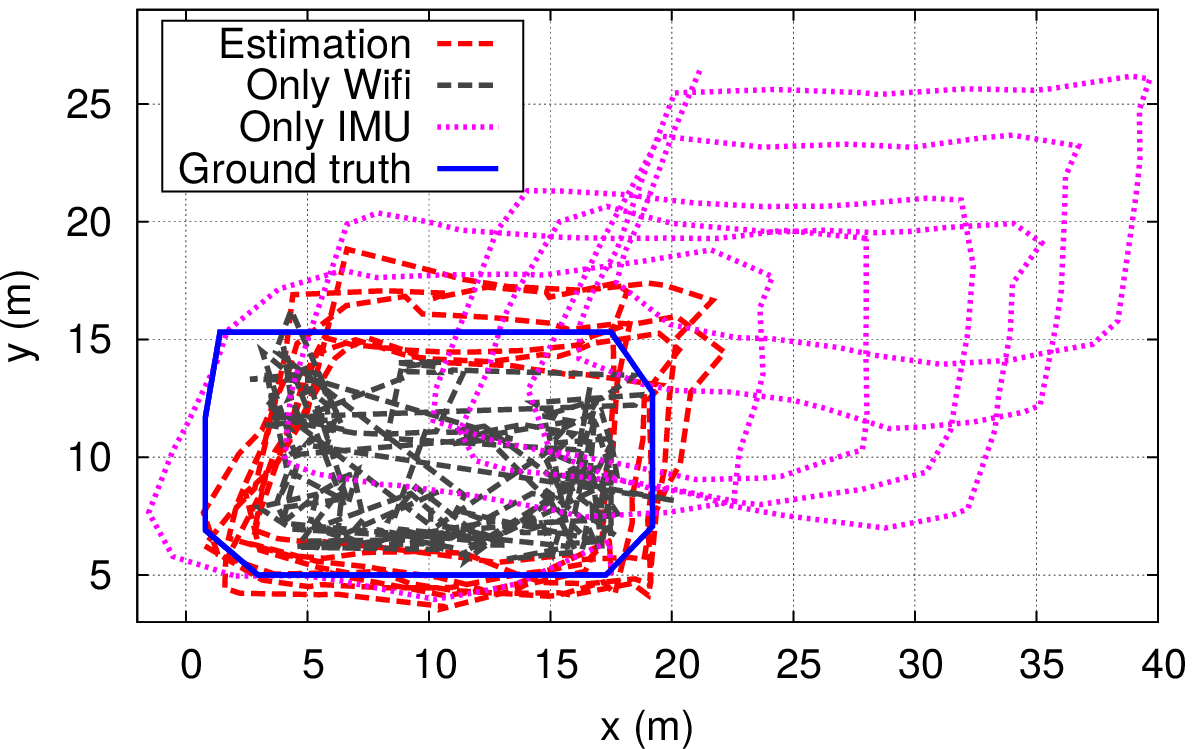}
        }
        \hspace{-0.3cm}
 \subfigure[Tracking error at different timestamps]{
\label{fig:tracking_error}
        \includegraphics[height=0.2105\textwidth]{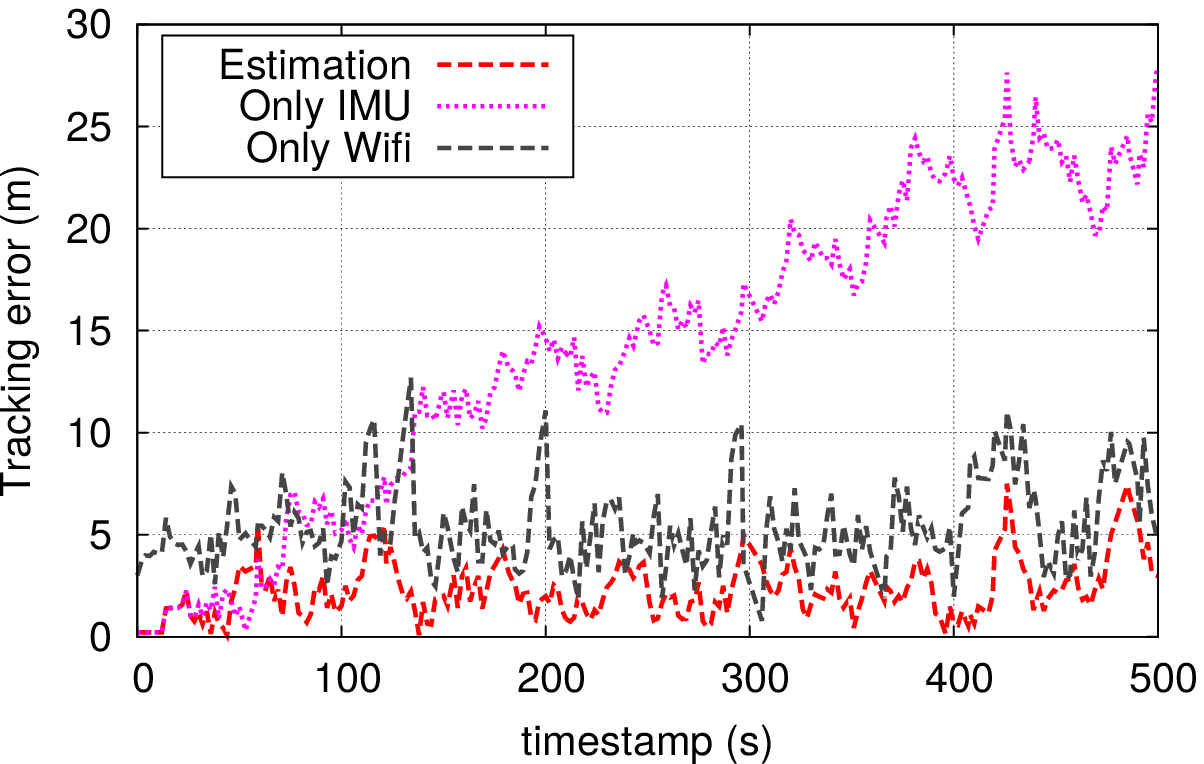}
        }
        \hspace{-0.3cm}
        \subfigure[Impact under different IMU noise scales]{
\label{fig:impact_of_different_imu_noise}
    \includegraphics[height=0.2105\textwidth]{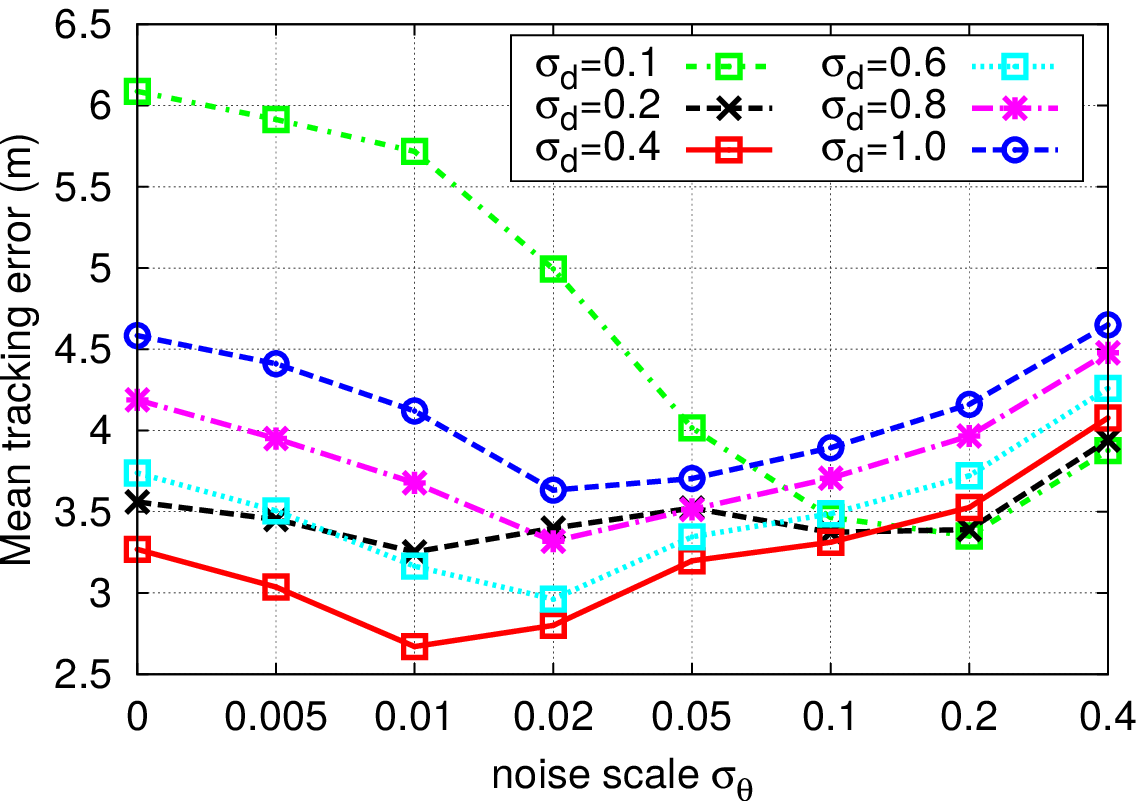}
    }    
   \caption[Tracking of fusion.]
{Performance evaluation. 
(a) Estimated trajectory with Wifi alone, IMU alone, combination of Wifi and IMU, and ground truth.
(b) Tracking error at different timestamps.
(c) Mean tracking accuracy under different scales of noise added to IMU.
}
\label{fig:example_track}
\vspace{-0.5cm}
\end{figure*}

%%%%%%%%%%%%%%%%%%%%%%%%%%%%%%%%%%%%%%%%%%%%%%%%%%%%%%%%%%%%%%%%%%%%%%%%%%%%%%%%%%%%%%%%%%%%%%%%%%%%%%%%
\section{Experimental Results}
\label{experimental_evaluations}
%%%%%%%%%%%%%%%%%%%%%%%%%%%%%%%%%%%%%%%%%%%%%%%%%%%%%%%%%%%%%%%%%%%%%%%%%%%%%%%%%%%%%%%%%%%%%%%%%%%%%%%%
%First, setting up AP and FM transmitters. 

%%%%%%%%%%%%%%%%%%%%%%%%%%%%%%%%%%%%%%%%%%%%%%%%%%%%%%%%%%%%%%%%%%%%%%%%%%%%%%%%%%%%%%%%%%%%%%%%%%%%%%%%
\subsection{Experimental Setups}
\label{experimental_setups}
%%%%%%%%%%%%%%%%%%%%%%%%%%%%%%%%%%%%%%%%%%%%%%%%%%%%%%%%%%%%%%%%%%%%%%%%%%%%%%%%%%%%%%%%%%%%%%%%%%%%%%%%
We evaluated our approach in an office environment with a size of 25\,m$\times$14\,m, as shown in Fig.\,\ref{fig:experimental_setup}. 
This environment consists of concrete walls, soft room partitions, furniture, and equipments.
Nine access points (ASUS RT-N12HP) are installed with known positions.
The phone processes the IMU data with a frequency of 50 Hz and sends the computed results $\U_t$ to a server once a step is detected. 
A Sony Z2 phone is used to retrieve the signal strength from the APs and upload them to the server with a frequency of 0.5 HZ . 
%The APs offer RSS values ranging from -70\,dBm to -30\,dBm. 
%The server used for sensor fusion is running on an Intel Core i5-4200M\,@\,2.5GHz CPU, with 4GB RAM. 

During our experiment, a user held a mobile phone (see Fig. \ref{fig:experimental_snapshot}) and walked along a rectangle path multiple times with a normal speed.
In total, he traveled approx. 648.2 meters in 831 seconds with an average velocity of 0.8 m/s. 
This resulted in a track consists of 415 Wifi and IMU measurements. 
%All data are send through through Wifi. 
To record the ground truth, we placed 302 visual landmarks on the walls. 
When the user passed by the landmarks, he is asked to press a button on the phone to send the ID of the landmark to the server.
The positions of these landmarks are measured before. 
%We define mean tracking error as the Root Mean Square Error (RMSE) between estimations and the ground truth.
%The mean tracking error is defined as the Root Mean Square Error (RMSE) between estimations and the ground truth.
A snapshot of the experiment is shown in Fig. \ref{fig:experimental_snapshot}. 

To compare to the traditional fingerprinting-based approach, we recorded the Wifi measurements manually at 41 locations as reference fingerprints as shown in Fig.\,\ref{fig:experimental_setup}. 
The locations of these positions are known before hand. 
At each reference position, we recorded Wifi measurements for 3 minutes. 
%Fig.\,\ref{fig:experimental_setup} showed the reference positions of the fingerprints we collected during the experiment.
We implemented a traditional fingerprinting-based approach based on the cosine similarity\,\cite{cos_similarity} and WKNN for a comparison. 

%Each experiment is repeated 40 times (estimations with the mapping extension are running almost every time to track effects of the uniform random  particle filter initialization),
%thus  giving  us  at  least 213 estimation samples over all 71 tags.
We use a grid-based representation to compute our sequence-based fingerprinting map. 
We discretize the environment into two-dimensional grids with a fixed grid size.
The location sequence in each grid is represented by the ranking of distance from the centroid of this grid to APs.
We performed various experiments to evaluate the performance our approach. 
%In all experiments, we conducted experiments on the same track. 

%In particular, Sect.\,\ref{experimental_localization_evaluation} evaluated the localization accuracy under different parameters for some test points 
%and Sect.\,\ref{tracking_experiment} presented the results of mobile user tracking.

%\textcolor{red}{Sensor model representation, grid-based representation.
%All of the data are sent via wifi network the server, which is responsible for the data fusion.}

%%%%%%%%%%%%%%%%%%%%%%%%%%%%%%%%%%%%%%%%%%%%%%%%%%%%%%%%%%%%%%%%%%%%%%%%%%%%%%%%%%%%%%%%%%%%%%%%%%%%%%%%
\subsection{{Tracking Performance With and Without Integrating IMU}}
\label{sect:fusion}
%%%%%%%%%%%%%%%%%%%%%%%%%%%%%%%%%%%%%%%%%%%%%%%%%%%%%%%%%%%%%%%%%%%%%%%%%%%%%%%%%%%%%%%%%%%%%%%%%%%%%%%%
We first examined the tracking accuracy with and without incorporating IMU. 
%Then we integrated the both measurements into the particle filter. 
%and the number of particles $N=1000$. 
We set the noise scale $\sigma_d=0.4$ and $\sigma_\theta=0.01$. 
We fix the grid size to $2.0$\,m and $\lambda=0.01$. The number of particles $N$ is set to 1000 and $k$ is fixed to be 4.
The tracking results are shown in Fig.\,\ref{fig:trajectory} and Fig.\,\ref{fig:tracking_error}. 
As can be seen from this figure, integrating IMU clearly gives a better result. 
For example, we obtained a mean tracking accuracy of $2.67$\,m by integrating IMU, 
which leads to an improvement of $48.8\%$, as compared to the result without IMU ($5.22$\,m). 
This is because IMU is precise to measure the change of position over short periods of time, 
therefore can be used to improve the overall tracking accuracy.
With IMU alone, the track will drift due to the accumulative characteristics.
For example, the mean tracking error of IMU alone is $27.5$\,m, while our approach achieves an accuracy of $2.67$\,m. 
%Due to its cumulative characteristic, the IMU will get worse for longer tracks and our approach is able to eliminate the cumulative error. 
%As compared with the final error of the IMU, our approach is much more precise.

We evaluated the tracking accuracy under the impact of different noise scales ($\sigma_d$ and $\sigma_\theta$) added to the IMU. 
We choose the number of particles $N=1000$ and the results are shown in Fig. \ref{fig:impact_of_different_imu_noise}. 
It can be seen from this figure, the best setting of parameters is $\sigma_d=0.4$ and $\sigma_\theta=0.01$.
A too large or too small noise scale obviously gives a bad result. 
%This can be seen from the rise of the curves for $\sigma_d>0.6$.
%This is because a too large noise scale will introduce too much noise to the IMU sensor and results in an unstable estimation, 
%thus giving a bad tracking result. 
%On the other hand, a too small $\sigma_d$ and $\sigma_\theta$ also lead to bad tracking results. 
%For example, for $\sigma_\theta=0$ or $0.1$, 
%the tracking result gets worse with the decrease of $\sigma_d$ for $\sigma_d<0.4$.
%This is because a too small $\sigma_d$ and $\sigma_\theta$ is not be able to capture the noise the from the IMU sensor, and may place a small
%number of particles around the true pose of the tracked target, which leads to poor tracking performance.

\begin{figure*}
  \centering
  \subfigure[Impact of different number of particles and parameter $\lambda$]{
\label{fig:number_particles}
        \includegraphics[height=0.22\textwidth]{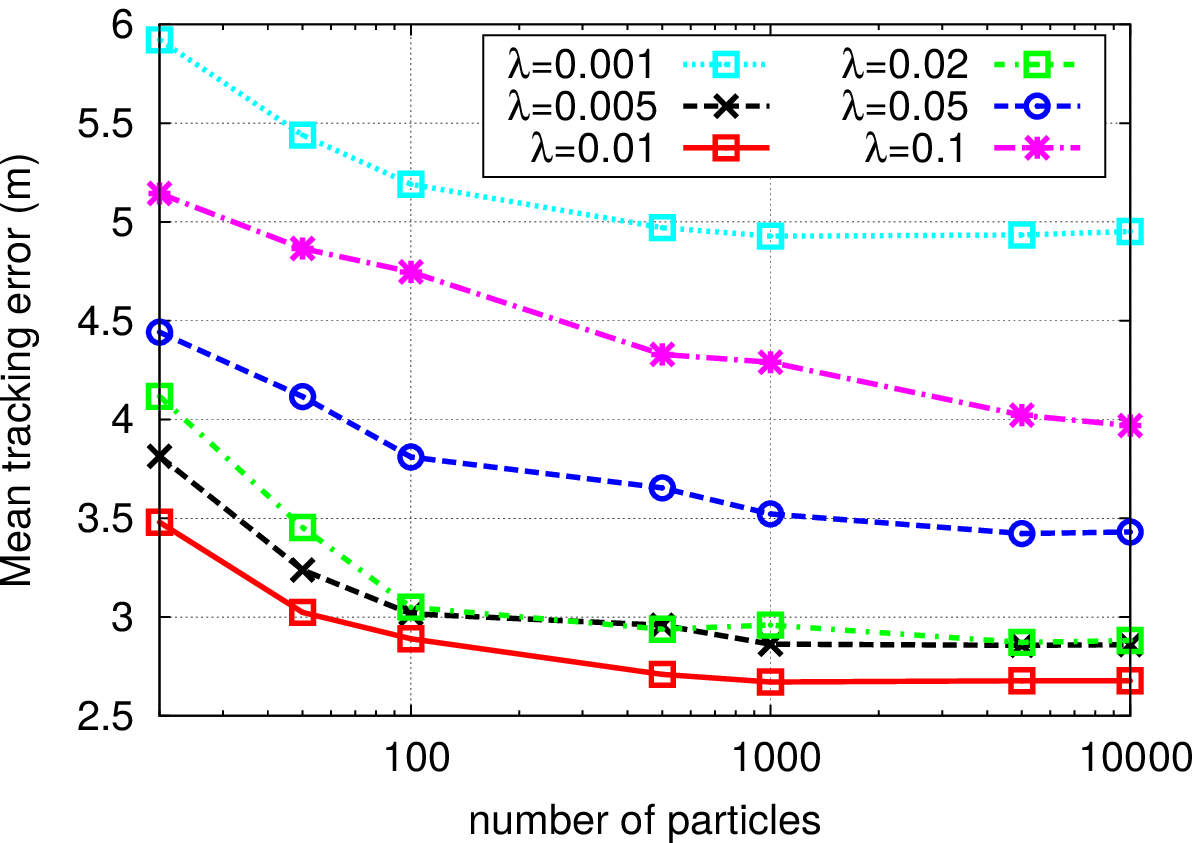}
        }        
    \subfigure[Impact under various step lengths]{
\label{fig:different_step_length}
    \includegraphics[height=0.22\textwidth]{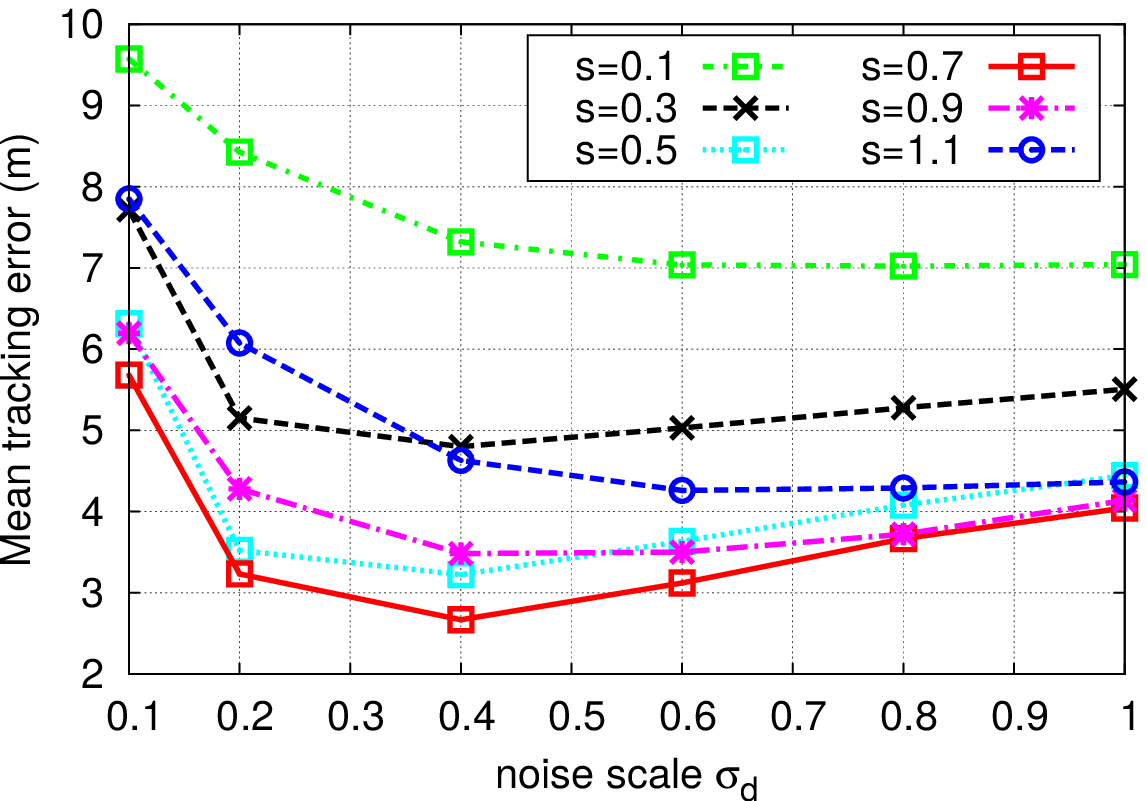}
    }        
  \subfigure[Compare to the traditional fingerprinting-based approach]{
\label{fig:compare_to_wknn}
        \includegraphics[height=0.22\textwidth]{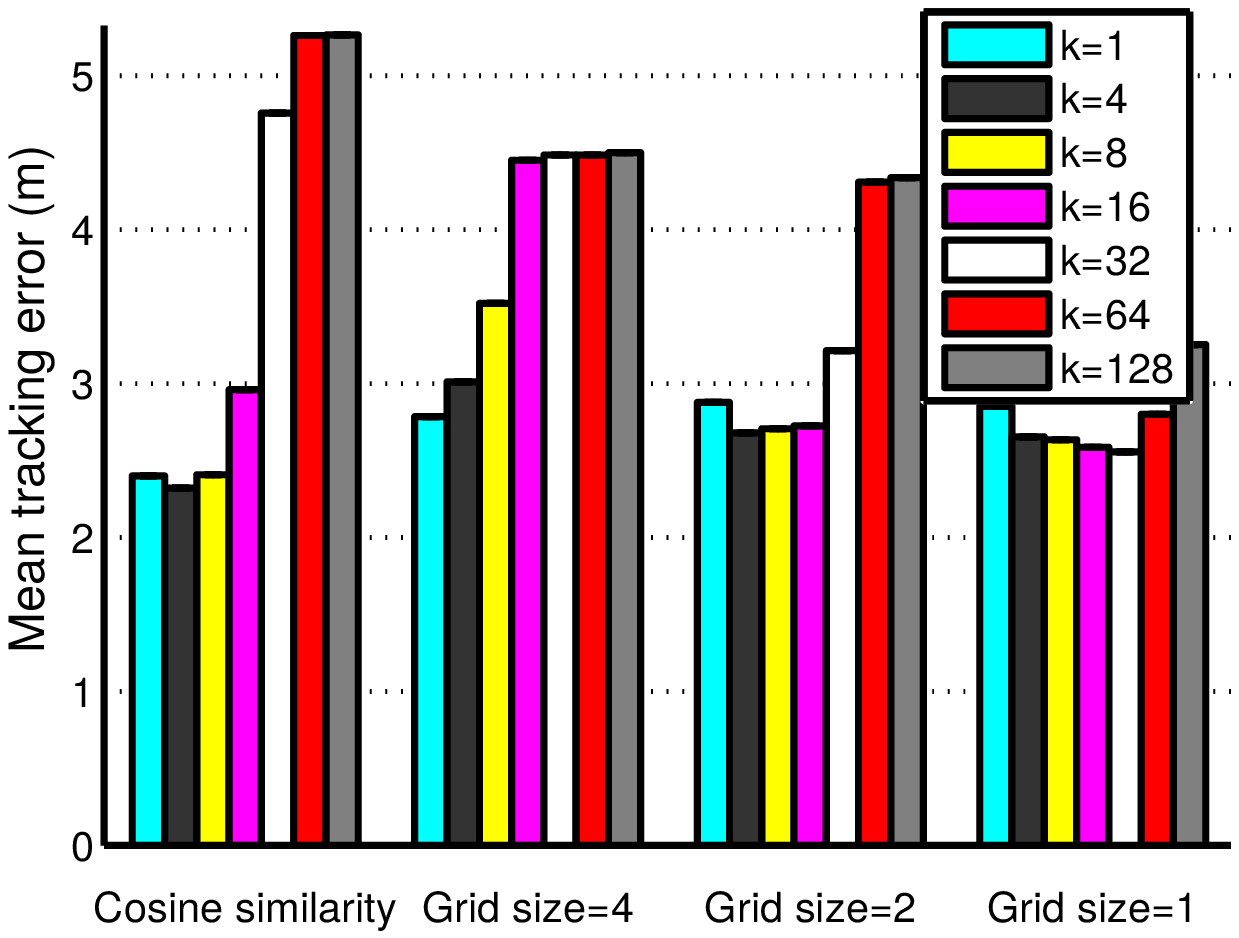}
        }
   \caption[Tracking of fusion.]
{Performance evaluation. 
(a) Mean tracking error under different number of particles and $\lambda$.
(b) Mean tracking error under the impact of different step lengths.
(c) Comparison of our approach to a traditional fingerprint-based approach with different number of nearest neighbors $k$.}
\label{fig:example_track}
\vspace{-0.5cm}
\end{figure*}

%But the Levenshtien distance metric deals with insertions and deletions of symbols, 
%which is taken into account the overall structure of the sequence and better capture the length of the symbols in each string is not the same. 
%Also, the Levenshtien distance metric has a well-bounded criterion for calculating the distance that offers higher accuracy compared to hamming distance. 

%%%%%%%%%%%%%%%%%%%%%%%%%%%%%%%%%%%%%%%%%%%%%%%%%%%%%%%%%%%%%%%%%%%%%%%%%%%%%%%%%%%%%%%%%%%%%%%%%%%%%%%%
\subsection{{Impact of Different Number of Particles}}
\label{sect:partilce_size}
%%%%%%%%%%%%%%%%%%%%%%%%%%%%%%%%%%%%%%%%%%%%%%%%%%%%%%%%%%%%%%%%%%%%%%%%%%%%%%%%%%%%%%%%%%%%%%%%%%%%%%%%
We examined the tracking accuracy under different number of particles $N$, as shown in Fig.\,\ref{fig:number_particles}. 
As can be seen from this figure, the tracking accuracy gets worse with smaller $N$ (e.g., $N\leq100$). 
%We observed that the robot could not navigate with smaller N than 20. 
With $N\geq1000$, we achieved nearly the same tracking accuracy. 
Obviously, the mean computational time required for larger $N$ increases due to the increasing number of particles.
Our experiments show that integrating one measurement with $N=1000$ on an Intel Core i5-4200M@2.50 GHz CPU with 4 GB RAM 
only requires 6.02 ms, which satisfies the requirement of real-time processing.
We also show the impact of $\lambda$ on the tracking performance in Fig.\,\ref{fig:number_particles}.
Our experiments revealed that $\lambda=0.01$ is the best choice for all settings of $N$.
A too larger or too smaller $\lambda$ obviously leads to a poor result.

%%%%%%%%%%%%%%%%%%%%%%%%%%%%%%%%%%%%%%%%%%%%%%%%%%%%%%%%%%%%%%%%%%%%%%%%%%%%%%%%%%%%%%%%%%%%%%%%%%%%%%%%
\subsection{{Impact of Different Step Length $s$}}
\label{sect:step_length}
%\subsection{Performance Evaluation}
%%%%%%%%%%%%%%%%%%%%%%%%%%%%%%%%%%%%%%%%%%%%%%%%%%%%%%%%%%%%%%%%%%%%%%%%%%%%%%%%%%%%%%%%%%%%%%%%%%%%%%%%
We examined the tracking accuracy under various step length $s$ in Fig.\,\ref{fig:different_step_length}. 
We also varied $\sigma_d$ to see its impact on the tracking accuracy, due to its high impact on the tracking performance.
%Since the tracking is related to the noise level $\sigma_d$ added.
As can be seen from Fig.\,\ref{fig:different_step_length}, the tracking accuracy gets worse with a too large or too small step length. 
A choice of $s=0.7$ gives the best tracking results.
In addition, $\sigma_d=0.4$ leads to the best tracking accuracy, which is consistent with the findings in Sect.\,\ref{sect:fusion}.
%A incorrect step length need more noise added in order to get a good tracking accuracy.
The step length may be different for various persons, an algorithm to estimate the step length can be found in\,\cite{step_length_estimation_2012}. 
%and will be addressed in our future work.
%We observed that the robot could not navigate with smaller N than 20. 

%%%%%%%%%%%%%%%%%%%%%%%%%%%%%%%%%%%%%%%%%%%%%%%%%%%%%%%%%%%%%%%%%%%%%%%%%%%%%%%%%%%%%%%%%%%%%%%%%%%%%%%
\subsection{{Compare to Traditional Fingerprinting-based Approach}}
\label{sect:compare_to_wknn}
%%%%%%%%%%%%%%%%%%%%%%%%%%%%%%%%%%%%%%%%%%%%%%%%%%%%%%%%%%%%%%%%%%%%%%%%%%%%%%%%%%%%%%%%%%%%%%%%%%%%%%%%

Finally, we compared our approach with a state-of-the art fingerprinting-based approach using cosine similarity and WKNN\,\cite{cos_similarity}. 
IMU information is integrated for both approaches with a noise setting of $\sigma_d=0.4$ and $\sigma_\theta=0.01$. 
We fix $\lambda=0.01$ and $N=1000$.
We choose different values of $k$ and various grid sizes of our approach to evaluate the tracking performance. 
The mean tracking accuracy is shown in Fig.\,\ref{fig:compare_to_wknn}. 
As can be seen from this figure, 
$k=4$ gives the best results for the traditional fingerprinting-based approach and our proposed approach with a grid size of 2.0\,m.
For both approaches, a too larger $k$ obviously leads to a worse result.
The traditional fingerprinting-based approach achieves a tracking accuracy of $2.35$\,m with $k=4$, 
which is slightly better than our sequence-based approach (with a mean tracking accuracy of $2.67$\,m).
Fingerprinting-based approach requires a phase to collect the measurements as the reference fingerprinting, which can be very time consuming. 
In contrast, our approach eliminates this time-consuming phase, and achieves comparable results, 
therefore may be considered as a good alternative to other existing state-of-the-art fingerprinting-based approaches.

In addition, it can be seen from Fig.\,\ref{fig:compare_to_wknn} that the optimal value of $k$ varies for different grid sizes.
To get a better accuracy, we need to assign a large $k$ for a small grid size.
For example, we achieve the best tracking accuracy with $k=4$ for a grid size of 2.0, 
while the best setting for a grid size of 1.0 is $k=32$.
Moreover, the tracking accuracy gets slightly better with a smaller grid size.
For example, the best accuracy achieved with a grid size of $1.0$ is 2.55\,m with $k=32$, 
which is an improvement of 4\% as compared with a grid size of 2.0\,m (i.e., 2.67\,m with $k=4$).
%Computing with a smaller grid size obviously consumes more time and more memory than a larger grid size, especially for large scale environments (see \cite{Liu_Relative_Globecom} in detail).
%The best accuracy achieved is better than a small grid size.

\begin{figure}
\centering
\includegraphics[width=0.40\textwidth]{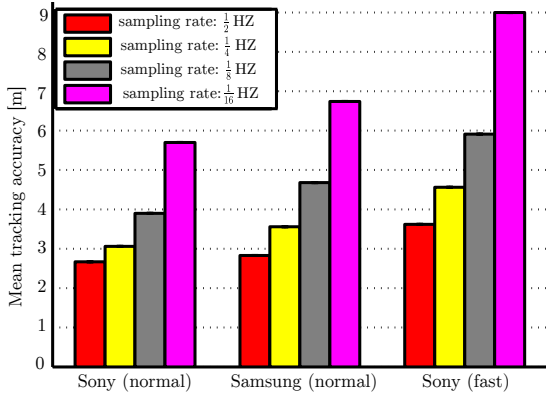}
\caption{Tracking accuracy of two different devices, various walking speeds, and different Wifi sampling rates.}
\label{fig:sampling_rate}
\vspace{-0.5cm}
\end{figure}

%%%%%%%%%%%%%%%%%%%%%%%%%%%%%%%%%%%%%%%%%%%%%%%%%%%%%%%%%%%%%%%%%%%%%%%%%%%%%%%%%%%%%%%%%%%%%%%%%%%%%%%%
\subsection{{Tracking Accuracy of Different Devices, Walking Speeds, and Sampling Rates}}
\label{sect:speed_device_diversity_walking_speed}
%\subsection{Performance Evaluation}
%%%%%%%%%%%%%%%%%%%%%%%%%%%%%%%%%%%%%%%%%%%%%%%%%%%%%%%%%%%%%%%%%%%%%%%%%%%%%%%%%%%%%%%%%%%%%%%%%%%%%%%%
We evaluated the tracking accuracy under different devices and different walking speeds using our sequence-based approach.
The same parameter setting in Sect.\,\ref{sect:compare_to_wknn} is applied for this series of experiments. 
We additionally recorded a track using the same Sony phone with a fast walking speed (approx. 1.0 m/s on average) and 
another track using a Samsung phone with a normal walking speed. 
We use a sampling rate of 0.5 HZ for the Wifi as previous. 
To compare the tracking accuracy under the impact of different Wifi sampling rates, 
we only integrate a part of Wifi measurements (i.e., all,  half, fourth, and eighth) on the recorded data, which is identical to a setting of different Wifi sampling rates ($\frac{1}{2}$Hz, $\frac{1}{4}$Hz, 
$\frac{1}{8}$HZ, and $\frac{1}{16}$HZ).
%We also show the sampling rate of the wifi to the tracking performance, so we took a wifi sample every 2 seconds (as default), 4 seconds, 8 seconds, and 16 seconds.
The tracking results are shown in Fig.\,\ref{fig:sampling_rate}. 
As can be seen from this figure, the two devices achieve similar tracking results (i.e., 2.67\,m for Sony phone and 2.85\,m for Samsung phone) with a normal walking speed.
In addition, a fast walking speed and a low sampling rate obviously lead to bad results, 
since in both cases there are not enough Wifi measurements to correct the IMU drift thus leading to poor tracking results. 

%%%%%%%%%%%%%%%%%%%%%%%%%%%%%%%%%%%%%%%%%%%%%%%%%%%%%%%%%%%%%%%%%%%%%%%%%%%%%%%%%%%%%%%%%%%%%%%%%%%%%%%%
\section{Conclusions}
\label{conclusions}
%%%%%%%%%%%%%%%%%%%%%%%%%%%%%%%%%%%%%%%%%%%%%%%%%%%%%%%%%%%%%%%%%%%%%%%%%%%%%%%%%%%%%%%%%%%%%%%%%%%%%%%%
In this paper,
we proposed a novel approach to combine similarity-based sequence technique and dead reckoning to localize and track users in indoor environments.
Our approach does not require any tedious site survey phase to construct the fingerprinting map, 
which is essential for the traditional fingerprinting-based approaches.
Extensive experiments were conducted to validate the performance of our approach.
We achieved a mean tracking accuracy of 2.67\,m, which is comparable to the traditional fingerprinting-based approach.
For the future work, we would like to evaluate our approach in large scale environments. 
In addition, we want to investigate a novel similarity measure to improve the tracking performance.

%%%%%%%%%%%%%%%%%%%%%%%%%%%%%%%%%%%%%%%%%%%%%%%%%%%%%%%%%%%%%%%%%%%%%%%%%%%%%%%%%%%%%%%%%%%%%%%%%%%%%%%%
%\section{Acknowledgment}
%\label{acknowledgment}
%%%%%%%%%%%%%%%%%%%%%%%%%%%%%%%%%%%%%%%%%%%%%%%%%%%%%%%%%%%%%%%%%%%%%%%%%%%%%%%%%%%%%%%%%%%%%%%%%%%%%%%%
%The authors would like to thank the support from Temasek Lab under Indoor Relative Positioning System project (No. IGDST1302024), National Science Foundation of China (No. 61550110244, 61601381, and 61471306), 
%and National Defense Scientific Research of China (No. B3120133002).

\ifCLASSOPTIONcaptionsoff
  \newpage
\fi



%\bibliographystyle{IEEEtran}

\bibliographystyle{IEEEtran}
\bibliography{literatur}

\end{document}